\documentclass[sigconf]{acmart}
\usepackage{multirow}
\usepackage{bbm}
\usepackage{subfigure}
\AtBeginDocument{%
  \providecommand\BibTeX{{%
    \normalfont B\kern-0.5em{\scshape i\kern-0.25em b}\kern-0.8em\TeX}}}

\copyrightyear{2022}
\acmYear{2022}
\setcopyright{rightsretained}
\acmConference[CODS-COMAD 2022]{5th Joint International Conference on Data Science & Management of Data (9th ACM IKDD CODS and 27th COMAD)}{January 8--10, 2022}{Bangalore, India}
\acmBooktitle{5th Joint International Conference on Data Science \& Management of Data (9th ACM IKDD CODS and 27th COMAD) (CODS-COMAD 2022), January 8--10, 2022, Bangalore, India}\acmDOI{10.1145/3493700.3493737}
\acmISBN{978-1-4503-8582-4/22/01}

\newcommand{\noteng}[1]{\textcolor{blue}{[\bf\small NG: #1]}}

\begin{document}

\title{An Application to Generate Style Guided Compatible Outfit} %SATCOGen}


 \author{$^*$Debopriyo Banerjee$^{1,2}$, $^*$Harsh Maheshwari$^1$, $^*$Lucky Dhakad$^1$, Arnab Bhattacharya$^1$, Niloy Ganguly$^2$, Muthusamy Chelliah$^1$, Suyash Agarwal$^1$}
 \authornote{First three authors contributed equally to this research.} 
\email{deb.ban89@gmail.com}
%  \author{Lucky Dhakad} 
%  \author{Arnab Bhattacharya} \author{Muthusamy Chelliah} \author{Suyash Agarwal}
\affiliation{%
   \institution{$^1$Flipkart Internet Pvt. Ltd., $^2$Indian Institute of Technology Kharagpur}
   \country{India}
   $^*$equal contributions}
\renewcommand{\shortauthors}{Banerjee and Maheshwari, et al.}
\begin{abstract}
  Fashion recommendation has witnessed a phenomenal growth of research, particularly in the domains of shop-the-look, context-aware outfit creation, personalizing outfit creation etc. Majority of the work in this area focuses on better understanding of the notion of complimentary relationship between lifestyle items. Quite recently, some works have realised that \emph{style} plays a vital role in fashion, especially in the understanding of compatibility learning and outfit creation. In this paper, we would like to present the end-to-end design of a methodology in which we aim to generate outfits guided by styles or themes using a novel style encoder network. We present an extensive analysis of different aspects of our method through various experiments. We also provide a demonstration api to showcase the ability of our work in generating outfits based on an anchor item and styles.
\end{abstract}







\maketitle

\section{Introduction}
Outfit recommendation is a relatively well studied area in which researchers aim to recommend outfits based on the notion of \emph{compatibility} between lifestyle items, see \cite{Vasileva:2018:LTAE,Song:2019:CompModelBook,Li:2020:CompositionalVisualCoherence,Wang:2021:GAttnNetVSE} for more details. A substantial volume of work has also been done on the specific area of personalised recommendations \cite{Zhan:2021:A3-FKG,Lu:2021:CVPR}. However, none of them specifically take style into account while learning compatibility within outfits. We realise that style is an essential component in modelling outfit compatibility as an outfit may look compatible under one style construct but not in another.

\begin{figure}[t]
    \centering
    \includegraphics[width=0.85\linewidth]{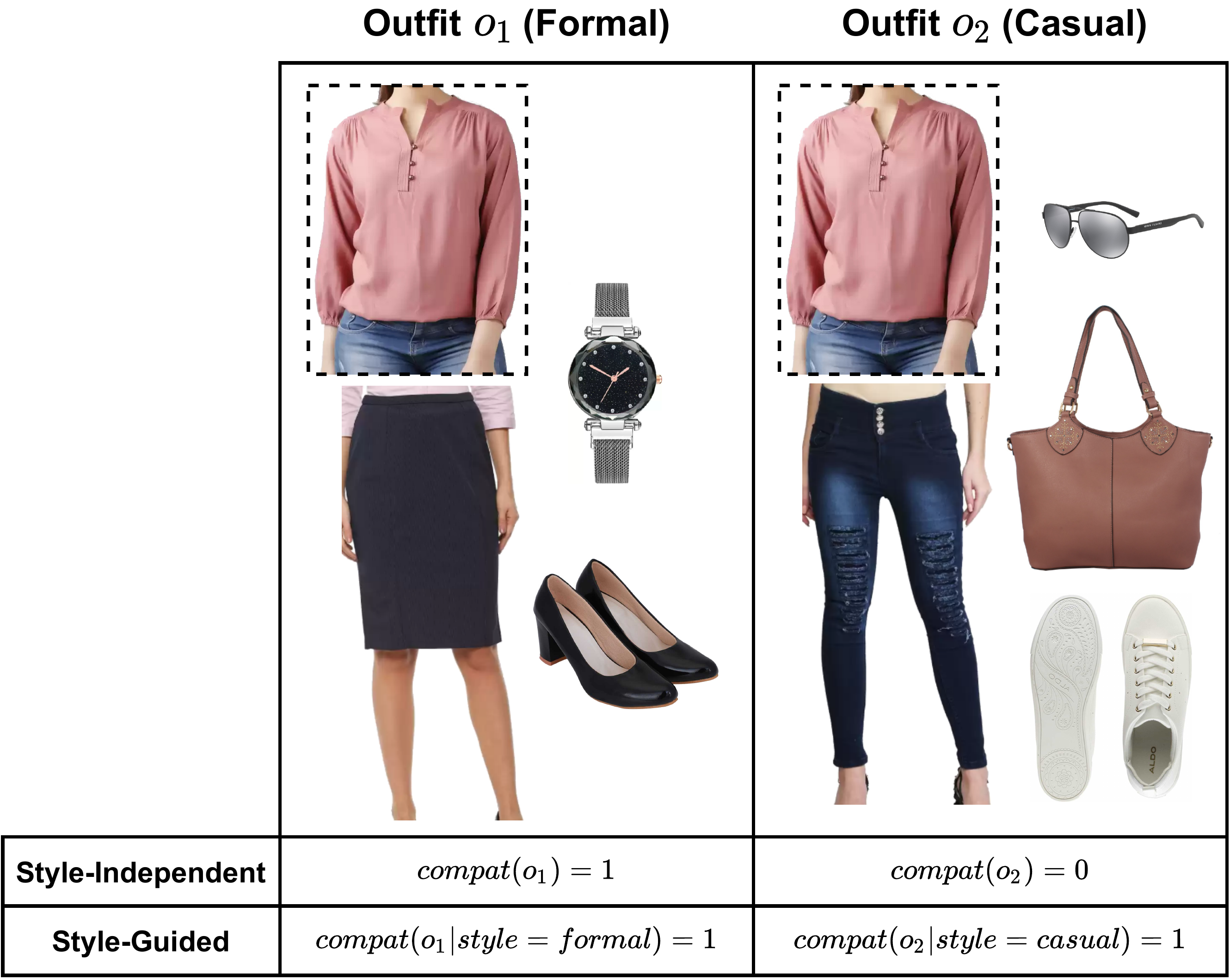}
    \caption{Illustration of the effectiveness of style-guided outfit generation over a style-independent variant. If \emph{formal} is the dominant style, the latter will only accept outfits from \emph{formal} style while rejecting from others. Style-guided methods, however, will accept outfits from multiple styles.}
     \vspace{-6mm}
    \label{fig:overview}
\end{figure}

An example of style guided outfit creation is provided in Figure~\ref{fig:overview}. The same top item  (highlighted by a dotted rectangle) is used to create two outfits under two different styles, namely \emph{Formal} and \emph{Casual}. This is useful in the situation where we consider that a user likes the top item but is doubtful about making the final purchase. A style-guided algorithm will have two advantages: (a) it will reject an outfit which may be otherwise compatible but not in accordance with the desired style, and (b) it will pick compatible outfits from different styles, hence expanding the choice to the user. A style-independent algorithm, on the other hand, gets biased towards the dominant style (assuming \emph{formal} style is dominant).

Style guided recommender system requires special attention from the research community as most of the work are unsupervised in nature. Some example research in this area are listed below:
%\begin{itemize}
    %\item 
    (a). Kuhn et al. \cite{kuhn:2019:Outfittery} refer an outfit as a style fit and do not explicitly use style information for modeling compatibility;
    %\item 
    (b). Jeon et al. \cite{Jeon:2021:FANCY} extract fashion attributes from full-body outfit images for classifying outfit style;
    %\item 
    (c). Li et al. \cite{li:2019:coherent} models outfit level style from item descriptions;
    %\item 
    (d). Singhal et al. \cite{Singhal:2020:VCP} models context and type jointly using Graph Neural Network (GNN), and style between item pairs are modeled using autoencoder without any explicit style information.
%\end{itemize}
Each of these works lack in one way or the other the ability to generate style guided outfits. 

Lai et al. proposed the Theme Matters paper~\cite{Lai:2020:ThemeMatters} (archived work) which comes closest to our work. It projects a supervised approach that applies theme-aware attention to item pairs with fine-grained category tags (e.g., long-skirt, mini-skirt, t-shirt, etc.). There are two specific drawbacks, the first of which is that such fine-grained category information is not always available and can be ambiguous if done manually. Secondly, the size of the model increases exponentially with the number of fine-grained categories.

We propose a {\bf Style-Attention-based Compatible Outfit Generation} (SATCOGen) framework that uses high-level categories (e.g., topwear, bottomwear, footwear, accessory, etc.) and outfit-level style information. It consists of a Style-Compatibility-Attention Network (SCA Net)~\cite{Lin:2020:AmazonFOCIR} and a novel style encoder network called Variational Style Encoder Network (VSEN) which encodes the style of an outfit into a latent space. This encoding is used to provide style-specific subspace attention, along with category information during the computation of embedding. Multiple loss functions ensure style encoding, general as well as style-specific compatibility. For the generation task, given an anchor item, beam search is used to generate style-specific outfit. We have provided a demonstration api to showcase the kind of outfits that are generated for an anchor item given various styles. 

\section{Methodology}
The fundamental philosophy guiding the work in this paper is that in practical circumstances compatibility between lifestyle items present within an outfit is contingent on the style to which the outfit belongs. In a nutshell, we make use of Style-Compatibility-Attention Network (SCA Net), a compatibility learning framework that makes use of features extracted from the image of an item based on category information as previously done by \cite{Lin:2020:AmazonFOCIR} and then add style component to it. The methodology of SATCOGen is explained in greater detail below.

A smooth latent vector representation for outfit style is learnt using Variational inference in a novel style encoder named \emph{Variational Style Encoder Network} (VSEN). There are two main trends in denoting an outfit,  as an ordered sequence of items \cite{Han:2017aa, Nakamura:2018:OutfitGenSTyleExtractionBiLSTMAE} or as \emph{set} \cite{bettaney2019aa, Chen:2019aa}. We choose the latter representation, which brings in two important properties, namely permutation invariance and allowance for varying length. This assumption enables us to select the \emph{set transformer} approach proposed in \cite{Lee:2019:SetTransformer} for our style encoder job. Keeping in mind that our work is not restricted only to compatibility learning  and also involves outfit generation, we ensure that every outfit style is represented by the first two moments of a Gaussian distribution which is proximal to the unit Gaussian $\mathcal{N}(0, \mathbbm{1})$, a mechanism we borrowed from Variational inference \cite{Blei:2017:VariationalInference}. This further ensures smooth representation of the latent style space. The advantage of this step will be clear during the outfit generation stage.

A vector, sampled from the Gaussian distribution representing the style of the outfit is used to classify the style of the outfit. This ensures that VSEN is able to capture specific information about the style of the outfit as well. Thus, given the styles and their corresponding style vectors, this module solves a multi-class classification problem using an MLP with $n$ layers. 

We modify the subspace attention network proposed in \cite{Lin:2020:AmazonFOCIR} to learn compatibility between items in an outfit. In the previous network, the image of an item within an outfit is passed through a ResNet18, and the embedding vector output is multiplied by learnt masks that help to learn the subspaces. The item and target categories are then consumed to estimate the subspace attention weights which subsequently leads to a weighted average of the masked embeddings to be denoted as the final embedding of the item in the tuple \texttt{<item, item category, target category>}. We tweak this and estimate the subspace attention weights by providing the outfit specific sample style vector from VSEN as an additional input. This helps to learn compatibility conditional on the style of the outfit.

There are four loss functions used in our method for learning style specific compatibility. We have the KL divergence loss from the VSEN network and the classification loss from the downstream job. We also have the compatibility loss from the SCA Net which is based on the popular triplet loss. And finally, we introduce one more loss function to account for penalisation when the wrong style is specified for an outfit. The overall loss for our method is given as the weighted sum of these four individual losses.
\begin{figure}[t]
    \centering
    \includegraphics[width=\linewidth]{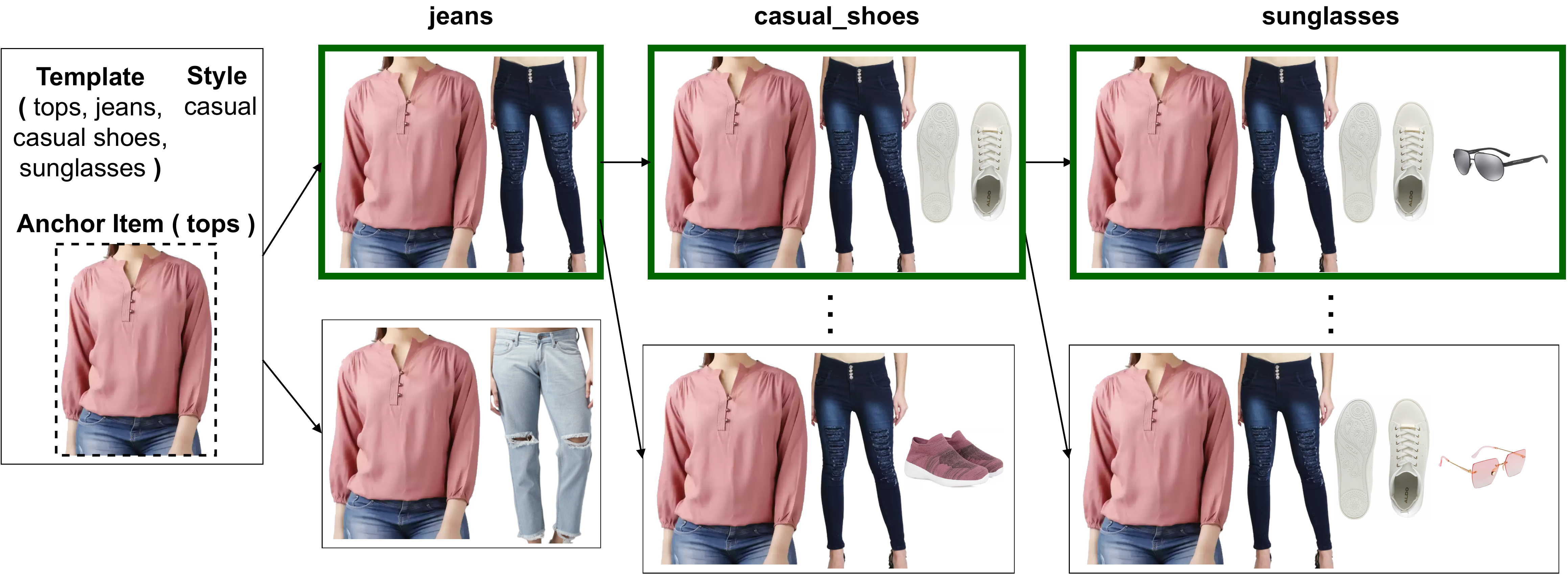}
    \caption{Beam search: Given a top-wear chosen by a user, and a template, the algorithm would go about generating outfits by sequentially adding items from each category in the template. 
    % For beam-width=2, the method first selects top-2 compatible \emph{jeans}, followed by top-2 each of \emph{casual shoes} and \emph{sunglasses}.
    }
    \vspace{-6mm}
    \label{fig:outfit_generation}
\end{figure}
\subsection{Outfit generation}\label{sec:outfit_gen}
A globally optimal outfit generation task is a non-trivial task since it is infeasible to look into all possible combinations. An approximate solution is provided in this case. First, embeddings are created for different target categories for an item and an associated style. Note that we know from the previous section that embedding computation requires us to provide a style vector for every item. If there is a reference outfit from the same style which we want to emulate, it is trivial to generate a style vector from VSEN using that outfit. However, in the absence of a reference outfit, there is no specific distribution to sample from. For this we pool the mean and variance of all outfits belonging to that specific style, and use a Gaussian with pooled moments. This distribution can be assumed to be representative of the style in question and enables us to generate a style vector from it.

Once we have estimated embeddings for each item, one can generate outfits based on the well known \emph{beam search method} \cite{Zhang:2020aa}, as is shown in Figure~\ref{fig:outfit_generation}.

% \begin{figure*}[t]
%     \centering
%     \includegraphics[width=\linewidth]{images/style-new.png}
%     \caption{Architecture of  Style-Attention-based Compatible Outfit generation framework (SATCOGen).}
%     \label{fig:proposed}
%     % \vspace{-4mm}
% \end{figure*}

% \begin{table}[ht]
% \caption{Distribution of compatible outfits across different styles.}
% \label{tab:outfit_stats}
% \centering
% \scalebox{1.0}{
% % \resizebox{\linewidth}{!}{
% \begin{tabular}{|c|c|c|c|}
% \hline
% % \textbf{Dataset} & \textbf{Style} & \textbf{Train} & \textbf{Val} & \textbf{Test}\\
% \textbf{Style} & \textbf{Train} & \textbf{Val} & \textbf{Test}\\
% \hline
% \if{0}
% \multirow{8}{*}{FK} & Party & 8183 & 1174 & 3018\\
% \cline{2-5}
% & Outdoor & 6280 & 1001 & 1937\\
% \cline{2-5}
% & Summer & 7061 & 1204 & 2551\\
% \cline{2-5}
% & Formal & 5136 & 840 & 1648\\
% \cline{2-5}
% & Athleisure & 16232 & 1981 & 2506\\
% \cline{2-5}
% & Winter & 16028 & 2135 & 4695\\
% \cline{2-5}
% & Casual & 5194 & 791 & 2034\\
% \cline{2-5}
% & Celebrity & 5424 & 808 & 1480\\
% \hline
% \fi
% % \multirow{7}{*}{Zal} & Work & 841 & 108 & 251 \\
% % \cline{2-5}
% Work & 841 & 108 & 251 \\
% \hline
%  Casual & 13062 & 1679 & 3917 \\
% \hline
%  Party & 1215 & 156 & 362\\
% \hline
%  Relax & 473 & 61 & 140\\
% \hline
%  Travel & 2128 & 272 & 631\\
% \hline
%  Athleisure & 1160 & 149 & 348\\
% \hline
%  Sporty & 534 & 68 & 160\\
% \hline
% \end{tabular}
% }
% \vspace{-4mm}
% \end{table}

\begin{table}[ht]
\caption{Distribution of compatible outfits across different styles.}
\label{tab:outfit_stats}
\centering
\scalebox{0.85}{
\begin{tabular}{|c|c|c|c|c|c|c|c|}
\hline
\textbf{Style}  & Work & Casual & Party & Relax & Travel & Athleisure & Sporty \\
\hline
\textbf{Train}  & 841 & 13062 & 1215 & 473  & 2128 & 1160 & 534 \\
\hline
\textbf{Val} & 108 & 1679 & 156 & 61 & 272  & 149 & 68 \\ 
\hline
\textbf{Test} & 251 & 3917 & 362 & 140  & 631 & 348 & 160 \\
\hline
\end{tabular}
}
\vspace{-2mm}
\end{table}

\begin{figure}[t]
    \centering
    % \subfigure[]{\label{subfig:annotation_interface_fk}
    % \includegraphics[width=\linewidth]{images/annotation_interface_fk.png}}
    % \subfigure[]{\label{subfig:annotation_interface_zal}
    \includegraphics[width=0.8\linewidth]{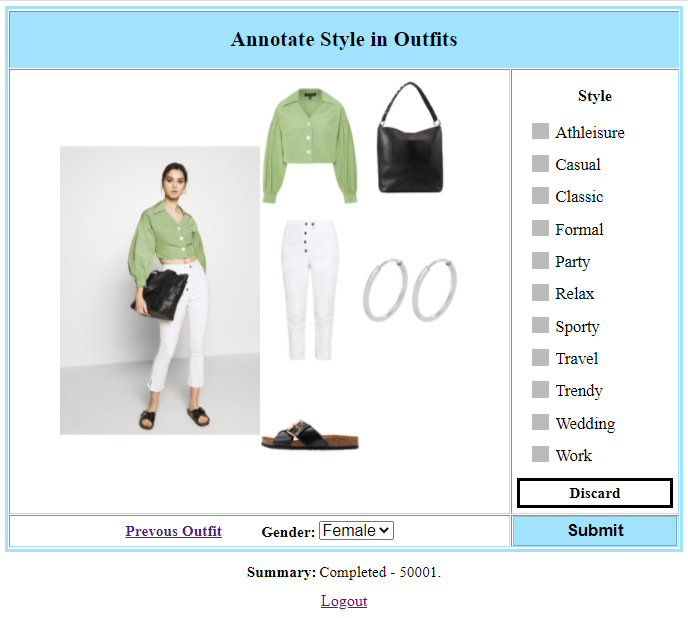}
    \caption{Annotation interface for curating the style of Zalando data. A complete outfit representing a \emph{look} from the website is provided along with individual items from the outfit. The annotator has to assign one or more appropriate style tags to the outfit, along with selecting the suitable gender for it. One can also discard the item if there is any form of discrepancy.}
    \vspace{-4mm}
    \label{fig:annotation_interface}
\end{figure}
\section{Dataset and Metrics}
We  created  a female outfit dataset with style annotations.\\
\if{0}
\textbf{Flipkart Dataset (FK):} This dataset consists of in-house fashion products available in Flipkart. Initially, five thousand outfits, consisting of multiple lifestyle items were manually curated with style tags, such as \emph{formal}, \emph{casual}, \emph{athleisure}, etc.
% \textcolor{red}{(AB: Why hare we referring to this image?)} 
However, since the number of curated outfits was insufficient to train a deep learning model, we used an efficient technique to expand the dataset. A similarity matching algorithm that makes use of item attributes and images as input was used to retrieve items that are visually similar to those present in a curated outfit and create new outfits, while also ensuring that the style of the original outfit is retained. %To confirm for compatibility of items within these created outfits, we take the help of human annotators. Each annotation has been done by more than one annotator and we consider only those data points where at least two agreed. This leaves us with a total of \textasciitilde{100K} outfits. The statistics of outfit distribution across different styles is given in Table~\ref{tab:outfit_stats}. 
We have six \noteng{table shows eight} higher level categories for items in our dataset. The distribution of outfits across various styles is given in table~\ref{tab:outfit_stats}.
\fi

\noindent \textbf{Zalando Dataset (Zal):} This dataset consists of items and outfits from the Zalando website\footnote{\url{https://www.zalando.co.uk/}}. In the website, \emph{outfit looks} are displayed with the option of shopping items from them. We merged semantically similar fine-grained item categories, which resulted in nine higher level categories. We scraped the looks and the corresponding set of items present in each of the looks. Since style tags are unavailable for majority of the looks, we recruited two human annotators to perform the task of annotating style tags in outfits. %Similar to the FK dataset, we collect two annotations for each outfit and consider data points that agree for both the annotators. 
We present the annotation interface in figure~\ref{fig:annotation_interface}, where we displayed an outfit look image along with the set individual items in the look on the left and center and a list of style options on the right. The annotator can assign one or more appropriate style tags to each outfit. We also provided a dropdown list to select the gender for which the outfit is primarily suitable. In case there is a mismatch in gender of the outfit look and the associated items or if there are any other discrepancies, the annotator has the option to select \emph{Discard}. Finally, there is an option to edit the immediately previous outfit and the number of completed annotations are shown below as a summary. It is ensured that each outfit is shown only once to each annotator.
% Finally, there is an option to edit the immediately previous outfit and the number of completed annotations are shown in summary. 

% This dataset consists of items and outfits from the Zalando website (\url{https://www.zalando.co.uk/}). In the website, \emph{outfit looks} are displayed with the option of shopping items from them. We merged semantically similar fine-grained item categories, which resulted in nine higher level categories. We scraped the looks and the corresponding set of items present in each of the looks. Since style tags are unavailable for majority of the looks, we recruit human annotators to perform the task of annotating style tags in outfits. %Similar to the FK dataset, we collect two annotations for each outfit and consider data points that agree for both the annotators. 
% We present the annotation interface in figure~\ref{fig:annotation_interface}, where we displayed an outfit look image along with the set individual items in the look on the left and center and a list of style options on the right. The annotator can assign one or more appropriate style tags to each outfit. We also provided a dropdown list to select the gender for which the outfit is primarily suitable. In case there is a mismatch in gender of the outfit look and the associated items or any other discrepancies, the annotator has the option to select discard. Finally, there is an option to edit the immediately previous outfit and the number of completed annotations are shown in summary. It is ensured that each outfit is shown only once to each annotator.

Even though an outfit can have multiple style tags, for this work we ensure that each outfit has a single style. In the situation that an outfit has different style tags from the two annotations, we choose the randomly select a style. After the annotation task, it was found that some of the styles were heavily under-represented. To mitigate this issue, we merged certain styles that are semantically similar and discarded some as well. For example, \textit{party} and \textit{wedding} were merged into \textit{party} while all outfits from \textit{classic} and \textit{trendy} were discarded. At the end we had \textasciitilde{28K} outfits. \\

% Even though an outfit can have multiple style tags, we consider a single style for each outfit in this paper. 
% For example, if an annotated outfit has three style tags \{ \textit{athleisure}, \textit{travel}, \textit{relax} \} in the first annotation and two style tags \{ \textit{travel}, \textit{relax} \} in the second annotation, then there is a common agreement of \{ \textit{travel}, \textit{relax} \} style tags for the outfit. Here we randomly choose either  \textit{travel} or \textit{relax} as the style tag. In case, the second annotation has two style tags \{ \textit{travel}, \textit{athleisure} \}, then we consider \textit{travel} as the style tag. 
% % \textcolor{red}{(AB: I did not understand last 3 sentences at all.)} 
% After the annotation job was over, it was found that some of the styles were heavily under-represented. To mitigate this issue, we merged certain styles that are semantically similar and discarded some as well. For example, \textit{party} and \textit{wedding} were merged into \textit{party} and \textit{formal} and \textit{work} to \textit{work}.  We discarded all outfits from the style tags \textit{classic} and \textit{trendy}. Finally, we had \textasciitilde{28K} outfits, where each item corresponds to one of nine higher level categories. \\

\noindent{\bf Metrics:} Two well known metrics are used to evaluate the performance of an outfit compatibility prediction model \cite{Han:2017aa, McAuley:2015:StylesSubstitutes}.\\ 
\noindent\textbf{Fill-in-the-blank Accuracy (FITB Acc.):} 
Given a set of items of an outfit with one missing item as query, the task is to predict the correct missing item from a list of four option items (where one is correct and three are incorrect) based on compatibility of each option item with the query set.\\ %For example, let us consider \{$x_{i,1}, x_{i,2}, \_, x_{i,4}$\} as the query $q_i$ corresponding to outfit $o_i$, where item $x_{i,3}$ is missing. The option items being \{$r_1, x_{i,3}, r_2, r_3$\}, where each $r_i$ ($i=1,2,3$) is an incorrect option. If the compatibility of the items in $q_i$ is maximum with $x_{i,3}$, then the prediction is correct, otherwise it is incorrect.
\textbf{Compatibility AUROC (Compat. AUC):} Given a set of positive and negative outfit samples, this metric helps in measuring the quality of predictions for compatible and incompatible outfits.

%Both are well known metrics used to evaluate the performance of an outfit compatibility prediction model \cite{Han:2017aa, McAuley:2015:StylesSubstitutes}. 
% The former is to identify the best item among the options in terms of compatibility with the query set, while the latter is to classify outfits into compatible and incompatible outfits. 
We constructed separate test sets for FITB and compatibility tasks similar to \cite{Vasileva:2018:LTAE} by creating soft negative (SN) and hard negative (HN) samples corresponding to each positive outfit sample. In case of SN, we sample random items from the matching higher level categories (e.g., topwear, bottomwear, footwear, etc.), whereas for HN, we do the sampling from matching fine-grained categories\footnote{We make use of fine-grained category information only during evaluation or testing phase.} (e.g., t-shirts, heels, shoes, etc.). It is to be noted that \emph{HN} samples are relatively harder to differentiate from positives than SN samples. This gradation helps in evaluating the performance of SATCOGen at various difficulty levels.
% For each outfit in the test data, we remove one product from the outfit and sample soft and hard negatives corresponding to its category to create the FITB test set. Compatibility test data is created similarly by sampling a mixture of negatives for each item in the outfit. 
We consider five FITB and Compatibility test datasets and report the mean performance. 
% Note that we have used the fine-grained category information only to measure model performance and not for training.

\section{Implementation Details}
The two modules of SATCOGen, namely VSEN and SCA Net make use of RestNet18 as the backbone CNN architecture. We freeze all the layers of ResNet18 except the last convolutional block, which is connected to a new fully connected layer that outputs a 64 dimensional embedding vector similar to the state-of-the-art compatibility learning methods \cite{Lin:2020:AmazonFOCIR, Vasileva:2018:LTAE}.

VSEN aggregates the CNN features of all the items in an outfit using the SAB Set Transformer \cite{Lee:2019:SetTransformer} with a hidden dimension of 32 and two heads  for mean and variance. The output of style vector of VSEN has a dimension of 64 which is followed by two fully connected MLP layers for the style classification task. Here we consider batch size as 128 and employ the Adam optimizer \cite{adam} with an initial learning rate of %$5 \times 10^{-5}$ for FK and 
$5 \times 10^{-6}$. % for Zal datasets. 
When combining the KL-Divergence loss with the cross entropy loss of classification, we consider 0.05 as the weight coefficient corresponding to KL-Divergence loss. 
% Note that, we have trained and frozen the VSEN module separately to be used in the rest of the network to expedite the process. 
% We get a style classification accuracy of 90.67\% and an F1 Score of 89.71\%.
After training the VSEN module, we freeze all of its parameters and use it in evaluation mode to train the rest of the SATCOGen framework. 
For learning the parameters of the SCA Net (five subspaces), we use the Adam optimizer with batch size of 32 triplets, and initial learning rate of %$1 \times 10^{-5}$ for FK and 
$1\times 10^{-6}$. % for Zal datasets. 
% The margin for $\mathcal{L}_{compat}$ and $\mathcal{L}_{stylecompat}$ was set to 0.3 and the weights were set to $1$ and $0.5$ respectively. 
The Attention Network of SCA Net transforms the concatenated one-hot-encoded category vectors and the extracted style vectors from VSEN into 32 dimensions, respectively (using a single fully connected layer), which are then forwarded to two fully connected layers (after concatenation). Finally, the output is the five subspace attention weights.

For outfit generation, we have optimized the inference code using native spark implementation to get item embedding given style and category information and to run beam search at scale on a large scale of volume of anchor and candidate items, resulting in 20X run time reduction compared to single box implementation. We compared the run time on ~60K anchor items with ~30K average child candidates per category, 5 average categories in beam search template and beam width of 3.
\begin{figure}[t]
    \centering
    \includegraphics[width=0.75\linewidth]{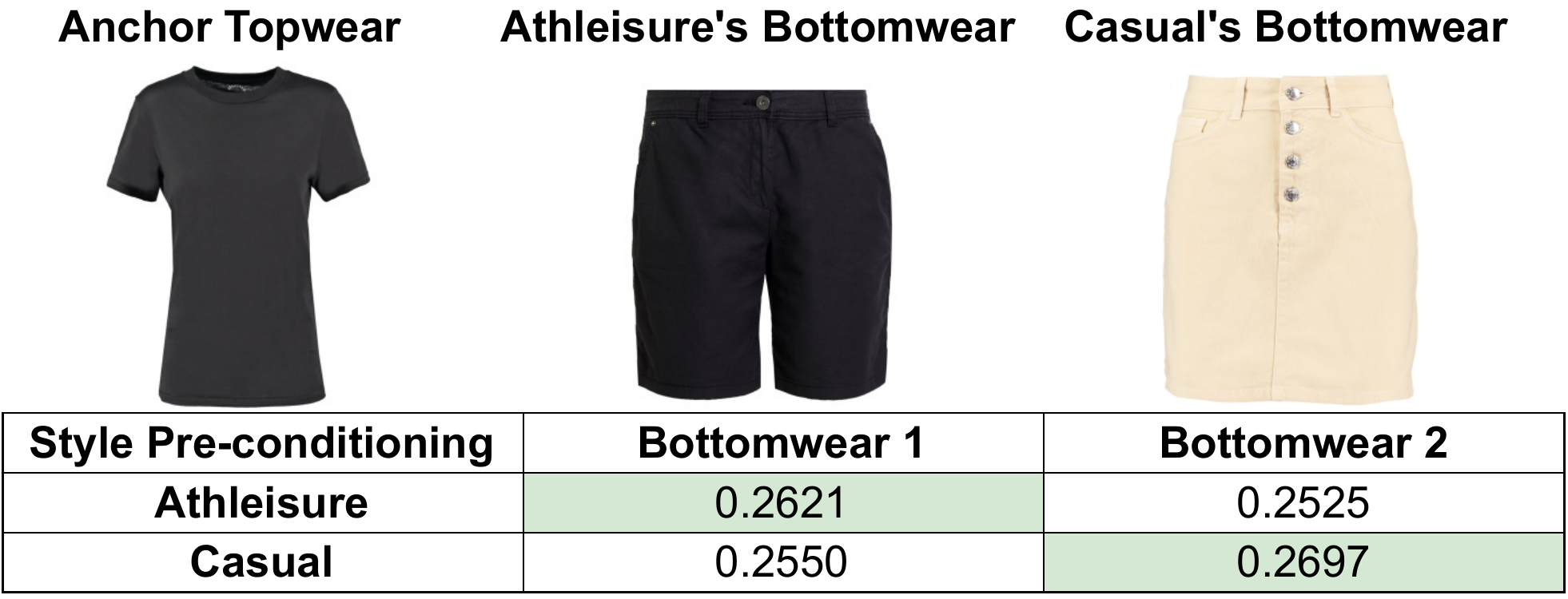}
    \caption{Demonstration of how SATCOGen is able to choose diverse style relevant bottomwears for a given parent topwear.
    }
    \vspace{-4mm}
    \label{fig:qualitative_results}
\end{figure}
\begin{table}[ht]
\caption{Comparison of compatibility learning for different methods on the Zalando dataset. We compute FITB and Compatibility ROC AUC with both hard and soft negatives.}
\label{tab:fitb-compatauc-metrics}
\centering
% \resizebox{\linewidth}{!}{  
\scalebox{0.9}{
\begin{tabular}{|c|l|c|c|}
\hline
\textbf{Method} & \textbf{Type}  & \textbf{FITB Acc.} & \textbf{Compat. AUC}\\
\hline
\multirow{2}{*}{TM} & SN & $47.79 \pm 0.07$ & $76.73 \pm 0.06$\\
\cline{2-4}
& HN & $43.78 \pm 0.25$ & $75.97 \pm 0.08$\\
\hline
\multirow{2}{*}{SATCOGen} & SN & $59.10 \pm 0.34$ & $88.58 \pm 0.08$\\
\cline{2-4}
& HN & $55.90  \pm 0.31$ & $86.96 \pm 0.06$ \\
\hline
\end{tabular}
}
\vspace{-4mm}
\end{table}
\section{Results}
% \noteng{Fig 4 is not referred, I think remove the baseline CSA-Net}
We show the efficacy of SATCOGen in generating superior outfits for online shopping portals by comparing its performance against Theme Matters (TM) \cite{Lai:2020:ThemeMatters} based on FITB Acc. and Compat. AUC scores. Table \ref{tab:fitb-compatauc-metrics} presents the FITB Acc. and Compat AUC results on SN and HN samples of Zal dataset for TM and SATCOGen (our proposed model). SATCOGen outperforms the TM results, mainly because of the requirement of fine-grained category information to have improved performance. Fig. \ref{fig:qualitative_results} shows an example of the quality of parent-child category combinations with style pre-conditioning. 

\section{Demonstration Interface}
\begin{figure}[h]
    \centering
    \includegraphics[width=\linewidth]{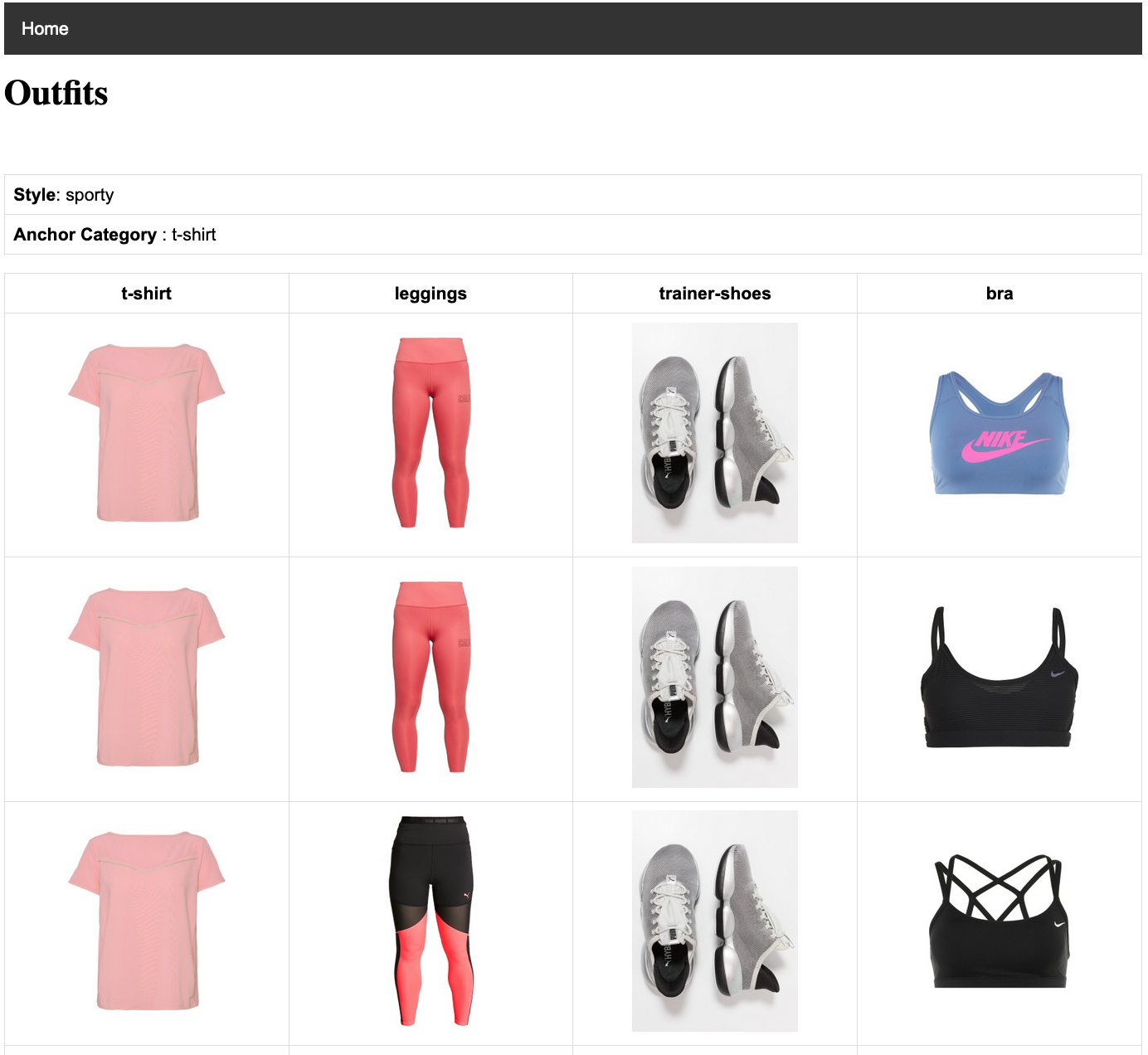}
    \caption{Screenshot of the Web Interface used for Demonstration.
    }
    \vspace{-4mm}
    \label{fig:demo}
\end{figure}

We have created a demonstration api to showcase the outfits generated by using SATCOGen framework given an anchor item, a style and candidate items from other categories. The outfit templates have been identified after discussion with fashion experts. An example of such template would be, \textit{('dress', 'heels', 'bag', 'jewellery')}. In the api, we provide users to select a category and subsequently an anchor item from the category for which the user wants to view outfits under different legitimate styles. 
The top-5 outfits per style are displayed for the selected anchor item. In figure-\ref{fig:demo}, we are showcasing the top-3 outfits for an anchor t-shirt given style \emph{sporty} and template \textit{'t-shirt', 'leggings', 'trainer-shoes', 'bra'}. The demo-api along with details and screenshots can be found here \footnote{\url{https://github.com/Lucky-Dhakad/SATCOGen-Demo-api}}.

\section{Conclusion}
In this paper, we presented SATCOGen~-~a novel outfit generation framework based on styles and evaluated its performance on Zal~-~a newly introduced outfit dataset with style tags associated with each outfit. In general, the outfit generation process using beam search algorithm is time consuming and not scalable for datasets having thousands of items for each category. We ensured scalability, by optimizing the code and making it suitable for execution on Hadoop Clusters, which reduced the execution time drastically. Finally, we presented a web-interface that demonstrates outfit generation starting from a chosen anchor item, a pre-defined style, and a template. In the future, we are planning to extend the outfit dataset with ethnic outfits (specific to Indian context) and   productionize the SATCOGen framework to provide an enhanced shopping experience to the customers. 
\bibliographystyle{ACM-Reference-Format}
\bibliography{ref}

% \appendix









\end{document}